\documentclass[preprint,1p]{elsarticle}
\usepackage[english]{babel}
\usepackage[utf8]{inputenc}
\usepackage{amsmath,amssymb}
\usepackage{booktabs,enumitem,multirow}
\usepackage{tikz}
\usetikzlibrary{automata}
\newcommand{\ncap}[1]{\gamma^{(#1)}}
\newcommand{\penratio}[1]{\sigma^{(#1)}}
\DeclareMathOperator{\Var}{Var}

\begin{document}

\title{A user subscription model in mobile radio access networks with network slicing
\tnoteref{t1}}
\author{José-Ramón~Vidal\corref{cor1}}
\ead{jrvidal@upv.es}
\author{Luis~Guijarro}
\ead{lguijar@upv.es}
\author{Vicent~Pla}
\ead{vpla@upv.es}
\cortext[cor1]{Corresponding author}
\address{Universitat Polit\`ecnica de Val\`encia, Camí de Vera s/n, 46021 Valencia, Spain}

\begin{abstract}
Network slicing is an architectural enabling technology 
that logically decouples the current cellular networks into
infrastructure providers (InPs) and Network Slice Tenants (NSTs). 
The network resources (e.g., radio access resources at each cell) are owned by the InP, 
and are shared by the NSTs to provide a service to their mobile users. 
In this context, we proposed a business model 
that includes resource allocation and user subscription to NSTs 
in a competitive setting, and provides, among other things, 
closed-form expressions for the subscription indicators in equilibrium of each NST at each cell. 
This model relies on the widely adopted logit model to characterize user subscriptions. 
However, as a consequence of user mobility and radio propagation, 
some of the underlying assumptions in the logit model do not hold. 
Therefore, further research is needed to assess the accuracy of the results 
provided by the logit model in a mobile radio scenario. 
We carry out a thorough evaluation of the validity of the model by comparing its results 
against those obtained through computer simulation. 
Our simulation model includes complete and realistic characterizations of user mobility 
and radio propagation. From the results, 
we conclude 
 in most cases the logit model provides valid results in a mobile radio scenario.
\end{abstract}

\date{February 26, 2023}
\maketitle

\section{Introduction}

Network slicing will play a crucial role in the 5G mobile network architecture  
by enabling service providers to offer a broad variety of services
with different characteristics
in an efficient and cost-effective manner. 
It is expected to provide the flexibility in network resource management 
needed to meet the stringent requirements of new services
in terms of performance, scalability and availability. 

A network slice is  a logical network that can be dynamically created
by the allocation of a collection of resources from a shared physical infrastructure.
In network slicing, virtualization of resources is a key concept 
that allows several operators to share an underlying physical infrastructure and make a more efficient use of the resources.

Within this context, in~\cite{guijarro2021} we envisioned a scenario where a physical network
owned by an InP is shared by a set of NSTs, and  network slicing is implemented
to request network resources as needed for service provision. 
The network consists of a set of resources or cells, 
where each of them includes all transmission resources associated to a cell
such as spectrum, radio equipment, etc.
In each cell there is a set of users who demand service. 
The network operation and maintenance is carried out by the InP. 
Each NSTs, acting as a network operator, has the right to a share of the network resources. 
This right may be derived from a number of different situations 
(e.g., each operator owned a network and they decided to pool their networks 
and outsource the operation of the pool to an InP).

A business model for the provision of service by NSTs to end users in this scenario
was proposed by the authors in~\cite{guijarro2021}.
Users can subscribe to any NST which provides service in the cell where they are located,
and the NSTs charge a price to their subscribers.
The NSTs dynamically request from the InP the access and core network amount of resources 
required to support the service.
 This business model was motivated as a more realistic
business model against previous works, such as~\cite{caballero2017b}, where each NST did not maximize its
profits, but an aggregate measure of its subscribers’ utility. The authors have, anyway,
undertaken other studies in this context, such as~\cite{sacoto2020}, and the most appropriate economic
model has been chosen in each study.

In order to slice the network and provide the service, the NSTs interact strategically.
These interactions were modeled as a game and the competition between the NSTs was analyzed.
In this game, the strategy of NSTs is a distribution of their share of the resources between the cells:
each NST assigns a weight to each cell with the condition that the sum of these weights
equals its share of the network resources.
A simple-form solution was proposed for the Nash equilibrium problem.
In the proposed solution the weight that an NST assigns to a cell 
is equal to the product of the share of the NST by the fraction of subscribers in the network that are in that cell.
The properties of this solution were characterized for a variety of cell capacities and user sensitivities.

In the aforementioned work, the problem of determining how many users subscribe to each NST at each cell
is addressed in a discrete choice framework,  
in which the choice is made from the set of NSTs and the decision to subscribe depends on the users' perceived utility, 
which, among other things, depends on the bit rate provided by each NST, 
which in turn depends on the resource capacity and on the resource management strategies put it in place by the NSTs.
The logit model provides a very good solution to model the utility of users
and has the enormous advantage of providing a closed-form expression for the choice probabilities.
The modeling of the user behavior through a logit model makes it  possible the formulation of the problem 
as a game and the study of its equilibrium.

The logit model belongs to a class of probabilistic choice consumer models
where the user utility of subscribing to an NST is the sum of two terms: 
an observed or objective part plus an unobserved or subjective part.
The unobserved part of the utility is modeled by means of a random variable with Gumbel distribution,
while the observed part of the utility is a deterministic value for each option.
In the business model described above, 
the observed part of the utility of a user depends on the bit rate provided by the NST the user subscribes to.
In~\cite{guijarro2021} we approximated this bit rate by the average bit rate provided by the NST,
which is a fixed value for all its subscribers, 
given the cell capacity, the weights chosen by all the NSTs in this cell, 
and the number of subscribers to the NST. 

However, if this business model is applied to a radio access mobile network, 
the actual bit rate provided by an NST to each of its subscribers in a cell is not a fixed value.
This has two causes: 
1) each user observes a different cell capacity depending on its location, 
and consequently for mobile users this capacity also varies in time;
2) the number of subscribers to an NST in a cell also varies in time, 
because mobile users change their subscription decisions over time,
and the bit rate that an NSTs can provide to each subscriber depends on its current number of subscribers.
Subscription changes are caused by handovers 
and also by users that may decide
to change their subscription decisions due to the variations in the bit rate provided by each NST in the current cell.

Due to this variability in the fixed part of the utility that users receive from each NST, 
the logit model cannot be applied directly to a scenario with mobile users.
A possible work-around to deal with this problem is to use the logit model as an approximation,
by assuming that 
in a given cell all subscribers to a given NST obtain exactly the same bit rate.
This constant bit rate should be calculated by considering that the cell capacity 
is a certain average of the actual cell capacity seen by all users in this cell over a sufficiently long period of time.
Similarly, it should also be considered that the number of subscribers to an NST in a cell 
is a certain average of the number of subscribers to that NST 
over a sufficiently long period of time.
Proceeding in this way amounts to considering that there is a random error 
between the value of the objective part of the utility (i.e., the bit rate) at each point in time 
and the constant value used to calculate the choice probabilities.

If the errors induced by the varying bit rate were independent, that is, 
if the error on the observed part of the utility for one subscription option was unrelated 
to the error for another option,
this error could be included in the unobserved part of the utility.
However, this does not happen here, since the errors in the the utility for the different options are highly correlated.
Then, the inclusion of the variation of the observed part of the utility 
in the variation of the unobserved part of the utility is just another approximation that uses the logit model.

Therefore, in this context the validity of the logit model could be compromised~\cite{shafiei2018}.
Even so, according to the literature~\cite{train2009},
the logit model might be able to capture the average utility fairly well even when
the observed part of the utility is random, 
since the logit formula seems to be fairly robust to misspecifications. 
This raises the question whether the subscription indicators
(subscription ratios at each cell and fraction of subscribers to each NST)
that are obtained using the logit model with the approximations described above
do accurately represent the corresponding averages of the actual subscription indicators.

To respond to this question, in this work we carry out a study using an extensive set of computer simulations
with  a versatile model which includes 
user mobility and a complete and realistic characterization of radio propagation.
The main subscription indicators are obtained by simulating a realistic setup in a scenario of network slicing 
with the business model described above.
Subsequently, these indicators are compared with those 
obtained from the mathematical results 
provided by the logit model under the assumptions made in~\cite{guijarro2021}.

The main contributions of this paper are the following:
\begin{enumerate}
\item We test the performance of a logit model of the user subscription decisions in a radio mobile scenario. 
By quantifying the error induced by the assumptions required by the logit model, 
we show that this model is robust enough to provide valid results in this scenario.
\item We generalize the validity of the results  in~\cite{guijarro2021}, 
obtained through an analysis based on a static model with simplistic assumptions on the radio channel capacity, 
to a scenario with mobility and a highly varying radio channel capacity because of radio propagation.
\item We identify the restrictions within which a logit model is appropriate 
to model the user behavior in a radio access mobile network.
\end{enumerate}

This paper is structured as follows. In Section~\ref{sec:relwork}, we conduct a review of the literature
regarding the use of the logit model in telecommunications economics. Section~\ref{sec:model} describes
the model for the InP, the NSTs, and the users. This model was presented in~\cite{guijarro2021}, but it
is summarized here to make this paper self-contained. Section~\ref{sec:problem} specifies the procedure
to obtain the results analytically from the model and by simulation, and the procedure
to compare them. In Section~\ref{sec:simmodel}, the simulation setup is described. In Section~\ref{sec:conclusion}, results
are presented and discussed. And finally, Section 7 draws the conclusions.

\section{Related work}\label{sec:relwork}

The discrete choice model framework was first proposed by McFadden~\cite{mcfadden1974}, 
and it has been widely adopted by disciplines such as marketing science (where the choices are advertised products; see e.g.~\cite{guadagni1983}) and transportation network studies (where the choices are modes or routes; 
see e.g.~\cite{small2007,meignan2007}). 
Among the available models, the logit model has been the most frequently adopted 
because it is both amenable to analysis and to econometric estimation~\cite{benakiva1985}. 
Some more recent works are the following: Reference~\cite{bortolomiol2022}, published
in 2022, uses the logit model as a discrete-choice model of demand in the context of
transportation network economics, in order to assess different market structures and
public policy actions. Reference~\cite{anderson2020}, published in 2021, uses the logit model within
the context of a comparative statics in an asymmetric oligopoly market structure where
competition takes place in both the short and the long run. And~\cite{motta2021}, published in 2021,
adopts the logit model for the market demand in an analysis of the impact of mergers in
a duopoly market structure where the firms compete in prices and investments.

The logit model in telecommunication economics was initially adopted as an econometric tool for regression analysis, 
e.g., the estimation of the demand of a second fixed telephone line and of mobile services 
in the US household market in 2000--2001~\cite{rodini2003}; 
or the impact of mobile number portability on the switching costs born by mobile subscribers in Spain 2001--2004~\cite{maicas2009}.

More recently, the use of the logit model as a descriptive model in analytical studies, 
as it is the case of this work and the previous one~\cite{guijarro2021}, 
has become widespread in network economics, as it was noted in~\cite{maille2014}. 
Some examples from the literature are~\cite{coucheney2013, caron2010}, 
where the price competition between ISPs is studied within the context of the net neutrality debate 
and the subscription decision is modeled using a logit model; \cite{shin2014}, 
where the study of consumer adoption behavior in the context of cloud computing services is modeled by a logit model; 
and~\cite{guijarro2017jsac}, where the user behavior is modeled as a logit model 
in the context of the competitive provision of sensor-based services.

However, there has been no discussion in the literature, to the best of our knowledge, 
about to which extent the logit model is accurate enough 
in a scenario as dynamic and uncertain  as that of mobile communications,
where the utility associated with each choice (here the subscription to one operator) 
depends on the radio channel characteristics and on the number of users in the cell, 
both of which are dynamic and uncertain.
This work aims at filling this gap in the literature. 

\section{Model}\label{sec:model}

\subsection{Resource allocation model}

A network is composed of a set $\mathcal{B}$ of access resources available to a set of users. 
In the case of a mobile service, 
each element of $\mathcal{B}$ represents the radio access resources in a cell in the network. 
From now on we will call the elements of $\mathcal{B}$ `cells', 
without any loss of generality. 
The cells are managed by an InP and leased by a set $\mathcal{S}$ of NSTs. 

The NSTs lease the radio resources in the cells to deliver service to the users. 
Each NST has the right to a share of the whole amount of resources. 
Let $s_i$ be the share of NST~$i$, with $\sum_{i \in \mathcal{S}} s_i= 1$. 

Each NST distributes its share of the resources between the cells as follows.
The  $i$-th~NST sets a weight $\omega_i^{(j)}>0$ for all $j \in  \mathcal{B}$, 
such that  $\sum_{j \in \mathcal{B}} \omega^{(j)}_i = s_i \leq 1$.
The weights set by all NSTs are notified to the InP.
Using that information the InP will assign to the set of subscribers to NST~$i$ at cell~$j$ an amount of resources given by
\begin{equation}\label{eq:rate}
R^{(j)}_i=\frac{\omega^{(j)}_i}{\sum_{t \in \mathcal{S}} \omega^{(j)}_t} c^{(j)},
\end{equation}
where $c^{(j)}$ is the capacity of cell~$j$, representing the total bit rate available at the cell.
By doing this, all available resources are allocated: 
$\sum_{i \in \mathcal{S}} R^{(j)}_i = c^{(j)}, \forall j \in \mathcal{B}$.

The above allocation scheme of the cell capacity among the subscribers is known as
``proportional-share algorithm''. The rationale of the above allocation rule draws from
the Fisher market model. The Fisher market is one of the most fundamental models
within mathematical economics. See~\cite{guijarro2021} for a detailed justification.

\subsection{Economic model}

Based on the resource allocation procedure described above, NSTs provide service to their subscribers 
and charge each subscriber a flat-rate price. Let $p_i$ be the price charged to NST~$i$ subscribers.

According to the logit model, given a discrete set of subscription options 
for a user in  cell $j$, the utility  of option~$i$,
which consists of subscribing to NST $i$, is 
\begin{equation}\label{eq:totalutility}
u^{(j)}_i=v^{(j)}_i + \kappa_{u,i},
\end{equation} 
where the term $v^{(j)}_i$ is the observed part of the utility,
while $\kappa_{u,i}$ is an unobserved value of the utility.
The observed part depends on the the bit rate obtained by the user when subscribing to NST $i$, 
and thus it is the same for all users receiving the same bit rate. 
The unobserved part has a different value for each pair user-NST, and 
they are modeled as independent and identically distributed random variables.

Following~\cite{maille2014}, the observed part of the utility of a user subscribed to NST~$i$ 
that receives a service in cell~$j$ with a bit rate of $r^{(j)}_i$ and pays a price $p_i$,  is modeled as
\begin{equation}\label{eq:utility}
v^{(j)}_i = \mu \log \left( r^{(j)}_i / p_i\right),
\end{equation}
where $\mu > 0$ is a sensitivity parameter. 

The dependence of the observed utility on the bit rate has been modeled by a positive logarithm, 
since there is evidence that service quality perception by telecommunication users follow logarithmic laws~\cite{reichl2011}. 
For the same reason, the dependence on the price is through a negative logarithm.
Accordingly, the bit rate-to-price ratio $r^{(j)}_i/p_i$ is the relevant magnitude for the observed utility. 

The unobserved user-specific part of the utility has been modeled 
assuming that each random variable $\kappa_{u,i}$ follows a Gumbel distribution of zero mean and parameter $\nu$~\cite{benakiva1985}.
 
We assume that all subscribers to a given NST in a cell receive the same amount of resources.
That is, the bit rate available to each user in cell~$j$ subscribed to NST~$i$ is 
\begin{equation}\label{eq:rate0price}
r^{(j)}_i=\frac{R^{(j)}_i}{n^{(j)}_i}, 
\qquad i \in \mathcal{S}, \quad j \in \mathcal{B},
\end{equation}
where $n^{(j)}_i$ is the number of subscribers to NST~$i$ in cell~$j$. 

From this utility model, and assuming that $c^{(j)}$ is a fixed value in cell~$j$,
so that all the users in cell~$j$ receive the bit rate given by~\eqref{eq:rate0price},
it can be shown~\cite{train2009} that the expected number of users in cell ~$j$ 
subscribed to NST~$i$ is given by
\begin{equation}\label{eq:users}
 \frac{n^{(j)}_i}{n^{(j)}} = \frac{(r^{(j)}_i/p_i)^\frac{\mu}{\nu}}{\sum_{t \in \mathcal{S}} (r^{(j)}_t/p_t)^\frac{\mu}{\nu}+(r_0^{(j)}/p_0)^\frac{\mu}{\nu}}, 
  \qquad i \in \mathcal{S}, \quad j \in \mathcal{B},
\end{equation}
where $n^{(j)}$ is the total number of users in cell~$j$, 
$\mu/\nu$ models the sensitivity of the users to the bit rate-to-price ratio, 
and $r_0^{(j)}/p_0=e^{v^{(j)}_0/\mu}$ with $v^{(j)}_0$  being the utility of not subscribing to any NST. 
The equal-price assumption is tantamount to assuming that the
competition is not in terms of prices but in terms of quality of service. A scenario where
the user subscription is driven by the currently received quality of service and not by the
price is not uncommon. As examples of this situation, we provide the following three~\cite{guijarro2021}:
(1) that a regulatory authority is enforcing price control over the provision of the service,
i.e., the service price is fixed by the authority and therefore not under the control of the
operators/tenants; (2) that the service price is agreed on a long-term contract, e.g., as
part of a bundled offer; and (3) that the operators/tenants are wholesale roaming service
providers, i.e., that they provide the service to roaming users.

If, in cell~$j$, a user would always be better off by subscribing than by not doing it,
then all users in cell~$j$ would subscribe. 
This case would be captured by letting $v^{(j)}_0 \rightarrow -\infty$ or, equivalently, 
setting $r_0^{(j)}=0$.
In practice, this would represent those cases where the utility of subscribing to certain NSTs 
is much greater than no subscribing to any NST: 
$r_i^{(j)}/p_i \gg r_0^{(j)}/p_0$ for some $i \in \mathcal{S}$.

It is assumed that the number of users in each cell is sufficiently high,
and therefore the users are price-takers.
Moreover, for simplicity, it is assumed the service prices charged by the NSTs are all the same: 
$p_i = p$ for all $i \in \mathcal{S}$. 
Also. the number of parameters can be reduced without loss of generality by setting $p_0=1$.   
Then,~\eqref{eq:users} can be rewritten as
\begin{equation}\label{eq:users1price}
 n^{(j)}_i = n^{(j)} \frac{ (r^{(j)}_i)^\frac{\mu}{\nu} }
 { \sum_{t \in \mathcal{S}}(r^{(j)}_t)^\frac{\mu}{\nu} + (p\,r_0^{(j)})^\frac{\mu}{\nu} }, 
\qquad i \in \mathcal{S}, \quad j \in \mathcal{B}.
\end{equation}

From~\eqref{eq:rate} and~\eqref{eq:rate0price}, the bit rate provided to NST~$i$ subscribers in cell~$j$ is
\begin{equation}\label{eq:rate1price}
r^{(j)}_i= \frac{\omega^{(j)}_i}{\sum_{t \in \mathcal{S}} \omega^{(j)}_t} \frac{c^{(j)}}{n^{(j)}_i}, 
\qquad i \in \mathcal{S}, \quad j \in \mathcal{B}.
\end{equation}

Let $\penratio{j}$ denote the \textit{subscription ratio} in cell $j$, 
defined as the fraction of users subscribed to an NST in cell~$j$:
\begin{equation}\label{eq:penetration}
\sigma^{(j)} \triangleq \frac{1}{n^{(j)}}\sum_{i \in \mathcal{S}} n^{(j)}_i,   \qquad j \in \mathcal{B};
\end{equation}
let $\ncap{j}$  denote the \textit{normalized capacity}, 
which represents the capacity per user and per monetary unit  
normalized by the \textit{virtual capacity} of the no-subscription option, $r_0^{(j)}$:
\begin{equation}\label{eq:normCapacity}
\ncap{j} \triangleq \frac{c^{(j)}}{n^{(j)} p\,r_0^{(j)}},   \qquad j \in \mathcal{B};
\end{equation}
and let $\rho_i^{(j)}$ denote the \textit{fraction of subscribers} in cell $j$ that subscribe to NST~$i$:
\begin{equation}\label{eq:rho_ij}
	\rho_i^{(j)} \triangleq \frac{n^{(j)}_i}{\sum_{t \in \mathcal{S}} n^{(j)}_t}
	             = \frac{n^{(j)}_i}{\penratio{j} n^{(j)}}.
\end{equation}

As shown in~\cite{guijarro2021}, if $r_0^{(j)}=0$, then $\penratio{j}=1$, and if $r_0^{(j)}>0$, 
then $\penratio{j}$ is the unique solution in $(0,1)$ of the equation
\begin{equation}\label{eq:penetration_NLeq}
      \penratio{j} - \left(\ncap{j}\right)^\beta
                     \frac{\sum_{t \in \mathcal{S}} \left(\omega^{(j)}_t\right)^{\beta}}
                          {\left(\sum_{t \in \mathcal{S}} \omega^{(j)}_t\right)^{\beta}}
                     \left(1-\penratio{j}\right)^{1-\beta} = 0,
      \end{equation}
where $\omega_i^{(j)},\; i\in\mathcal{S},\; j\in\mathcal{B}$ are the weights set by NST~$i$ at cell~$j$ and 
\begin{equation}\label{eq:beta}
\beta \triangleq \frac{\mu}{\mu+\nu}<1.
\end{equation} 

Finally,
the fraction of subscribers in cell $j$ that subscribe to NST~$i$ can be expressed
as a function of only the weights of cell~$j$~\cite{guijarro2021}:
  \begin{equation}\label{eq:n_rs}
     	 \rho_i^{(j)} =  
	         \frac{(\omega^{(j)}_i)^\beta}{\sum_{t \in \mathcal{S}}(\omega^{(j)}_t)^\beta}.
  \end{equation}

\section{Problem statement}\label{sec:problem}
To obtain the subscription indicators from the expressions in the previous section,
a fixed capacity $c^{(j)}$ and a constant number of users $n^{(j)}$ at each cell~$j$ 
have to be assumed.
This has been implemented by applying these expressions taking
a capacity of each cell fixed and equal to its spatial average capacity,
and a number of users in each cell fixed and equal to its time average number of users.
In order to verify if the subscription indicators 
obtained analytically under these assumptions
do accurately represent the corresponding averages of the actual subscription indicators,
they have been compared with the results obtained by simulation.
The simulations are conducted in a realistic situation
in which users observe a varying cell capacity as they move
and the number of users in each cell varies due to handovers
and changes in their subscription choice.

The relevant indicators to compare are the subscription ratio at each cell $\sigma^{(j)}$
and the fraction of subscribers to each NST at each cell $\rho_i^{(j)}$.  
From this comparison we intend to determine the accuracy of the  model of Section~\ref{sec:model} 
to capture the subscription behavior in a radio access mobile network.

\subsection{Results obtained by simulation}\label{sec:problemSim}

Let $\hat{\sigma}^{(j)}$ be the estimated subscription ratio at cell~$j$, 
and $\hat{\rho}_i^{(j)}$ the estimated fraction of subscribers to NST~$i$ 
at cell~$j$, both obtained by simulation.
To calculate these values, the time average number of subscribers 
to each NST at each cell must be estimated over the simulated time.
At the current simulation (virtual) time, let $\texttt{n}_i^{(j)}$ 
be the number of subscribers to NST~$i$ at cell~$j$,
and $\texttt{n}_0^{(j)}$ the number of users at cell~$j$ not subscribed to any NST,
and let $\hat{\texttt{n}}_i^{(j)}$ be the value of the estimated time average of $\texttt{n}_i^{(j)}$.
Every time a user at cell~$j$ subscribes or unsubscribes, 
both the correspondent $\texttt{n}_i^{(j)}$ and $\hat{\texttt{n}}_i^{(j)}$ are updated.

Let $\texttt{t}_k^{(j)}$ be the simulation time of the $k$-th subscription 
or unsubscription event at cell~$j$, 
assuming that the simulation time starts at $t=0$.
If the event at $\texttt{t}_k^{(j)}$ involves NST~$i$, $\hat{\texttt{n}}_i^{(j)}$ is updated through
\begin{equation}
\hat{\texttt{n}}_i^{(j)}  \leftarrow \frac{\texttt{t}_{k-1}^{(j)}\hat{\texttt{n}}_i^{(j)}+(\texttt{t}_k^{(j)}-\texttt{t}_{k-1}^{(j)})\texttt{n}_i^{(j)}}{\texttt{t}_k^{(j)}},
\end{equation} 
and $\hat{\texttt{n}}_0^{(j)}$ is updated alike. 
Next, $\texttt{n}_i^{(j)}$ and $\texttt{n}_0^{(j)}$ are updated as $\texttt{n}_i^{(j)}\leftarrow \texttt{n}_i^{(j)}+1$ 
and $\texttt{n}_0^{(j)}\leftarrow \texttt{n}_0^{(j)}-1$, if the user subscribes,
or $\texttt{n}_i^{(j)}\leftarrow \texttt{n}_i^{(j)}-1$ and $\texttt{n}_0^{(j)}\leftarrow \texttt{n}_0^{(j)}+1$, 
if the user unsubscribes.

At the end of the simulation, the estimated time average number of users at cell~$j$ is
\begin{equation} \label{eq:meanNj}
\hat{n}^{(j)}=\sum_{i \in \mathcal{S}}\hat{\texttt{n}}_i^{(j)}+\hat{\texttt{n}}_0^{(j)},  \qquad j \in \mathcal{B},
\end{equation}
and
\begin{equation}
\hat{\sigma}^{(j)} = \frac{\hat{n}^{(j)}- \hat{\texttt{n}}_0^{(j)}}{\hat{n}^{(j)}},   \qquad j \in \mathcal{B}
\end{equation}
and
\begin{equation}
\hat{\rho}_i^{(j)} =  
	         \frac{\hat{\texttt{n}}^{(j)}_i}{\hat{\sigma}^{(j)} \hat{n}^{(j)}}, 
\qquad i \in \mathcal{S}, \quad j \in \mathcal{B}.
\end{equation}

\subsection{Results obtained analytically}\label{sec:problemAnalytic}

The subscription ratios $\sigma^{(j)}, \; j \in \mathcal{B}$, are
calculated by solving~\eqref{eq:penetration_NLeq}, 
where the normalized capacity $\gamma^{(j)}$ is an estimate given by
\begin{equation}\label{eq:gammaEst}
\hat{\gamma}^{(j)}=\frac{\check{c}^{(j)}}{\hat{n}^{(j)} p\,r_0^{(j)}},   \qquad j \in \mathcal{B},
\end{equation}
where $\hat{n}^{(j)}$ is the estimated time average of the number of users at cell~$j$ obtained in the simulation~\eqref{eq:meanNj},
and $\check{c}^{(j)}$ is an estimate of a statistic of the capacity of cell~$j$.
The values of $\check{c}^{(j)}$ used in~\eqref{eq:gammaEst} are obtained from  the cell capacities perceived 
by the users in cell~$j$ at the subscription times.

Two sets of results have been calculated analytically: a first set where $\check{c}^{(j)}$ is the mean of the cell capacity; 
and a second set where $\check{c}^{(j)}$ is the median of the cell capacity.
Let $\bar{c}^{(j)}$ be the estimated mean of cell~$j$ capacity and $\hat{c}^{(j)}$ its estimated median.
Both $\bar{c}^{(j)}$ and $\hat{c}^{(j)}$ are estimated from a large set of measures made at random locations 
in the simulation setup described in Section~\ref{sec:simmodel},
which is justified because in the Random Waypoint Model all
cell locations have the same probability of being visited by users.
Let $\sigma^{(j)}(\bar{c})$ be the subscription ratio at cell~$j$ obtained analytically when $\check{c}^{(j)}=\bar{c}^{(j)}$ in~\eqref{eq:gammaEst};
and let $\sigma^{(j)}(\hat{c})$ be the subscription ratio at cell~$j$ obtained analytically when $\check{c}^{(j)}=\hat{c}^{(j)}$.
Note that the values of the fractions of subscribers obtained analytically, 
$\rho_i^{(j)}$ are the same in both sets because they are obtained directly from~\eqref{eq:n_rs}, and do not depend on the cell capacity.

Additionally, a third set of results has been calculated analytically, 
in order to evaluate a modification of the logit model
that includes the variation of the observed part of the utility 
in the variation of the unobserved part of the utility.
In the modified model, 
the cell capacity is decomposed into the sum of a fixed part equal to its mean plus a variable part of zero mean.
The observed part of the utility is computed from the fixed part of the capacity,
while the effect of the variable part of the capacity is added to the variable that models the unobserved part of the utility.
Note that, for a given user, the variations on the utility caused by variations on the capacity are the same regardless of the NST, 
so the independence assumption required to be included in the unobserved part in the logit model is not met here.
Then, this approach has to be understood as an approximation to the logit model whose validity needs to be evaluated. 

To obtain this third set of results, a modified value of $\beta$ is used in~\eqref{eq:penetration_NLeq} and~\eqref{eq:n_rs}.
Let $\tilde{\beta}$ denote the modified $\beta$;
it has to be such that the variable part of the capacity is added to the variable 
that models the unobserved part of the utility $\kappa_{u,i}$.
Combining~\eqref{eq:rate1price} and~\eqref{eq:utility} with~\eqref{eq:totalutility}, the total utility is
\begin{align}
\nonumber 
u^{(j)}_i & = 
\mu 
\left[
\log\frac{ \omega^{(j)}_i}{\sum_{t \in \mathcal{S}} \omega^{(j)}_t} 
+\log(c^{(j)})-\log(n^{(j)}_i)-\log(p) 
\right]  \\
& + \kappa_{u,i},
\end{align} 
and its variance, assuming that the variation on $n^{(j)}_i$ is negligible, is
\begin{equation}
\Var[u^{(j)}_i]= 
\mu^2\Var[\log(c^{(j)})]+ \Var[\kappa_{u,i}].
\end{equation} 

Since $\kappa_{u,i}$ follows a Gumbel distribution of zero mean and parameter $\nu$,
\begin{equation}
\Var[\kappa_{u,i}]=\frac{\pi^2}{6\nu^2}.
\end{equation} 
Then, in order to increase the variance of the unobserved utility an amount equal to the variance 
of the observed utility caused by the variance of the cell capacity, 
the unobserved utility is model with a new variable $\tilde{\kappa}_{u,i}$ with variance
\begin{equation}
\Var[\tilde{\kappa}_{u,i}]=\frac{\pi^2}{6\tilde{\nu}^2}=
\mu^2\Var[\log(c^{(j)})]+ \frac{\pi^2}{6\nu^2};
\end{equation} 
and the Gumbel parameter $\nu$ becomes
\begin{equation}\label{eq:alfaModified}
\tilde{\nu}=\frac{\nu}
{\sqrt
{1+6\left(\frac{\mu\nu}{\pi}\right)^2\Var[\log(c^{(j)})]}};
\end{equation}
and the user sensitivity parameter $\tilde{\beta}$~\eqref{eq:beta} becomes
\begin{equation}
\tilde{\beta}=\frac{\mu}{\mu+\tilde{\nu}}.
\end{equation} 

Let $\sigma^{(j)}(\tilde{\beta})$ and $\rho_i^{(j)}(\tilde{\beta})$ denote, respectively, 
the subscription ratios at cell~$j$ and the fractions of subscribers at cell~$j$ 
obtained analytically using $\tilde{\beta}$.

\section{Simulation model}\label{sec:simmodel}

In this Section the simulation setup is described. 
In Section~\ref{sec:networkmodel} the mobile network configuration, 
the channel propagation model and the user mobility model are described. 
The actions performed by the users to evaluate the utility of each of their subscription options 
and to update their subscriptions are described in Section~\ref{sec:usermodel}.

\subsection{Network model}\label{sec:networkmodel}

\begin{figure}[htb]
\begin{center}
\resizebox{.7\columnwidth}{!}{
\begin{tikzpicture}
\def\s{0.886}; 
\def\t{0.5};\def\y{0};\foreach \x/\z in{-\s/44,\s/43}{
\draw (\x+\s,\y+\t) -- (\x,\y+2*\t)-- (\x-\s,\y+\t)-- (\x-\s,\y-\t)-- (\x,\y-2*\t)-- (\x+\s,\y-\t)-- cycle;
\draw(\x,\y)node{\z};}
\def\y{-3*\t};\foreach \x/\z in{-4*\s/29,-2*\s/28,0/45,2*\s/26,4*\s/25}{
\draw (\x+\s,\y+\t) -- (\x,\y+2*\t)-- (\x-\s,\y+\t)-- (\x-\s,\y-\t)-- (\x,\y-2*\t)-- (\x+\s,\y-\t)-- cycle;
\draw(\x,\y)node{\z};}
\def\y{-6*\t};\foreach \x/\z in{-7*\s/47,-5*\s/46,-3*\s/30,-1*\s/8,1*\s/7,3*\s/27,5*\s/41,7*\s/40}{
\draw (\x+\s,\y+\t) -- (\x,\y+2*\t)-- (\x-\s,\y+\t)-- (\x-\s,\y-\t)-- (\x,\y-2*\t)-- (\x+\s,\y-\t)-- cycle;
\draw(\x,\y)node{\z};}
\def\y{-9*\t};\foreach \x/\z in{-6*\s/48,-4*\s/11,-2*\s/10,0/9,2*\s/5,4*\s/4,6*\s/42}{
\draw (\x+\s,\y+\t) -- (\x,\y+2*\t)-- (\x-\s,\y+\t)-- (\x-\s,\y-\t)-- (\x,\y-2*\t)-- (\x+\s,\y-\t)-- cycle;
\draw(\x,\y)node{\z};}
\def\y{-12*\t};\foreach \x/\z in{-7*\s/32,-5*\s/31,-3*\s/12,-1*\s/2,1*\s/1,3*\s/6,5*\s/23,7*\s/22}{
\draw (\x+\s,\y+\t) -- (\x,\y+2*\t)-- (\x-\s,\y+\t)-- (\x-\s,\y-\t)-- (\x,\y-2*\t)-- (\x+\s,\y-\t)-- cycle;
\draw(\x,\y)node{\z};}
\def\y{-15*\t};\foreach \x/\z in{-6*\s/33,-4*\s/14,-2*\s/13,0/3,2*\s/20,4*\s/19,6*\s/24}{
\draw (\x+\s,\y+\t) -- (\x,\y+2*\t)-- (\x-\s,\y+\t)-- (\x-\s,\y-\t)-- (\x,\y-2*\t)-- (\x+\s,\y-\t)-- cycle;
\draw(\x,\y)node{\z};}
\def\y{-18*\t};\foreach \x/\z in{-7*\s/50,-5*\s/49,-3*\s/15,-1*\s/17,1*\s/16,3*\s/21,5*\s/56,7*\s/55}{
\draw (\x+\s,\y+\t) -- (\x,\y+2*\t)-- (\x-\s,\y+\t)-- (\x-\s,\y-\t)-- (\x,\y-2*\t)-- (\x+\s,\y-\t)-- cycle;
\draw(\x,\y)node{\z};}
\def\y{-21*\t};\foreach \x/\z in{-6*\s/51,-4*\s/35,-2*\s/34,0/18,2*\s/38,4*\s/37,6*\s/57}{
\draw (\x+\s,\y+\t) -- (\x,\y+2*\t)-- (\x-\s,\y+\t)-- (\x-\s,\y-\t)-- (\x,\y-2*\t)-- (\x+\s,\y-\t)-- cycle;
\draw(\x,\y)node{\z};}
\def\y{-24*\t};\foreach \x/\z in{-3*\s/36,-1*\s/53,1*\s/52,3*\s/39}{
\draw (\x+\s,\y+\t) -- (\x,\y+2*\t)-- (\x-\s,\y+\t)-- (\x-\s,\y-\t)-- (\x,\y-2*\t)-- (\x+\s,\y-\t)-- cycle;
\draw(\x,\y)node{\z};}
\def\y{-27*\t};\foreach \x/\z in{0/54}{
\draw (\x+\s,\y+\t) -- (\x,\y+2*\t)-- (\x-\s,\y+\t)-- (\x-\s,\y-\t)-- (\x,\y-2*\t)-- (\x+\s,\y-\t)-- cycle;
\draw(\x,\y)node{\z};}
\def\x{0}\foreach \y in {-\t,-7*\t,-13*\t,-19*\t,-25*\t}{
\draw[->,red,very thick](\x,\y)--(\x,\y-\t);
\draw[->,red,very thick](\x,\y)--(\x+\s*\t,\y+0.5*\t);
\draw[->,red,very thick](\x,\y)--(\x-\s*\t,\y+0.5*\t);}
\foreach \x in {-3*\s,3*\s}{
\foreach \y in {-4*\t,-10*\t,-16*\t,-22*\t}{
\draw[->,red,very thick](\x,\y)--(\x,\y-\t);
\draw[->,red,very thick](\x,\y)--(\x+\s*\t,\y+0.5*\t);
\draw[->,red,very thick](\x,\y)--(\x-\s*\t,\y+0.5*\t);}}
\foreach \x in {-6*\s,6*\s}{\foreach \y in {-7*\t,-13*\t,-19*\t}{
\draw[->,red,very thick](\x,\y)--(\x,\y-\t);
\draw[->,red,very thick](\x,\y)--(\x+\s*\t,\y+0.5*\t);
\draw[->,red,very thick](\x,\y)--(\x-\s*\t,\y+0.5*\t);}}
\end{tikzpicture}}
\caption{Simulated urban micro-cell grid}\label{fig:cellgrid}
\end{center}
\end{figure}
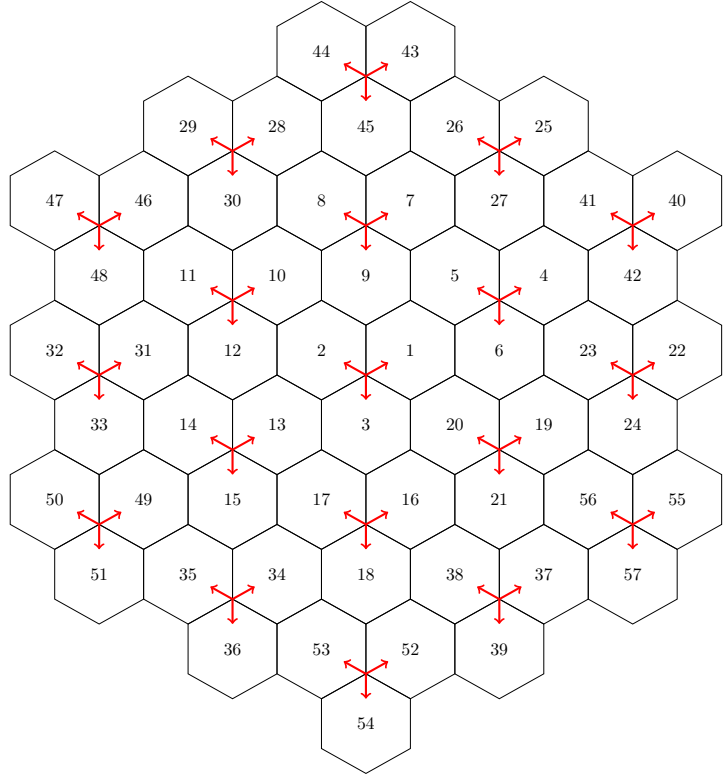

The simulation setup is based on the IMT-Advanced evaluation guide-lines for urban micro-cell deployment scenario \cite{itu2009}, 
consisting in a cell grid made up of 19 clusters of 3 cells each.
Therefore the network consists of $|\mathcal{B}|=57$ cells as shown in Fig.~\ref{fig:cellgrid}.

The simulations are conducted in a wrap-around configuration of this grid
in such a way that users who leave the grid re-enter it from the opposite side. 
For example, a user leaving cell 40 through its eastern boundary 
enters cell 48 through its western boundary,
a user leaving cell 36 through its western boundary 
enters cell 43 through its eastern boundary,  and so on.

Base stations are located at the center of each cluster of three hexagonal cells.
Inter station distance is $\text{ISD}= 200$\,m. 
At each base station, a three-sector antenna provides coverage to the three surrounding cells, as shown in Fig.~\ref{fig:cellgrid}.
Each antenna sector transmits a power of $P_0=41$\,dBm and has a gain, which 
depends on the user position, given by
\begin{equation}\label{eq:antennaGain}
G= G_{\text{max}} - \min\left\{ 12 \left(\frac{\theta}{\theta_{\text{3dB}}}\right)^2, A_m\right\} \; \mbox{dB},
\end{equation}
where $G_{\text{max}}=17$\,dB is the maximum antenna gain, 
$\theta$ is the angle of the antenna sector beam with the user position vector relative to the base station,
$\theta_{3\text{dB}}=70 \pi / 180$\,rads is the $3$\,dB beam width,
and $A_m=20$\,dB is the maximum gain loss.

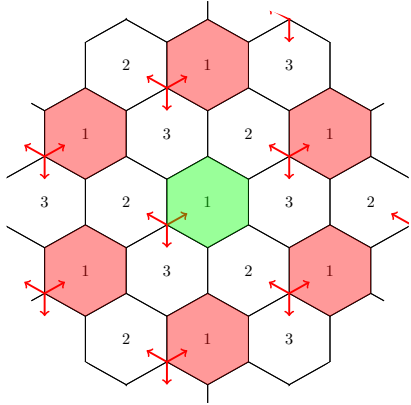
\begin{figure}[htb]
\begin{center}
\resizebox{0.4\columnwidth}{!}{
\begin{tikzpicture}
\def\s{0.886}; \def\t{0.5};
\clip (\s,-12*\t) circle(8.8*\t); 
\def\y{0};\foreach \x/\z in{-\s/2,\s/1}{
\draw (\x+\s,\y+\t) -- (\x,\y+2*\t)-- (\x-\s,\y+\t)-- (\x-\s,\y-\t)-- (\x,\y-2*\t)-- (\x+\s,\y-\t)-- cycle;
\draw(\x,\y)node{\z};}
\def\y{-3*\t};\foreach \x/\z in{-4*\s/2,-2*\s/1,0/3,2*\s/2,4*\s/1}{
\draw (\x+\s,\y+\t) -- (\x,\y+2*\t)-- (\x-\s,\y+\t)-- (\x-\s,\y-\t)-- (\x,\y-2*\t)-- (\x+\s,\y-\t)-- cycle;
\draw(\x,\y)node{\z};}
\def\y{-6*\t};\foreach \x/\z in{-7*\s/2,-5*\s/1,-3*\s/3,-1*\s/2,1*\s/1,3*\s/3,5*\s/2,7*\s/1}{
\draw (\x+\s,\y+\t) -- (\x,\y+2*\t)-- (\x-\s,\y+\t)-- (\x-\s,\y-\t)-- (\x,\y-2*\t)-- (\x+\s,\y-\t)-- cycle;
\draw(\x,\y)node{\z};}
\def\y{-9*\t};\foreach \x/\z in{-6*\s/3,-4*\s/2,-2*\s/1,0/3,2*\s/2,4*\s/1,6*\s/3}{
\draw (\x+\s,\y+\t) -- (\x,\y+2*\t)-- (\x-\s,\y+\t)-- (\x-\s,\y-\t)-- (\x,\y-2*\t)-- (\x+\s,\y-\t)-- cycle;
\draw(\x,\y)node{\z};}
\def\y{-12*\t};\foreach \x/\z in{-7*\s/2,-5*\s/1,-3*\s/3,-1*\s/2,1*\s/1,3*\s/3,5*\s/2,7*\s/1}{
\draw (\x+\s,\y+\t) -- (\x,\y+2*\t)-- (\x-\s,\y+\t)-- (\x-\s,\y-\t)-- (\x,\y-2*\t)-- (\x+\s,\y-\t)-- cycle;
\draw(\x,\y)node{\z};}
\def\y{-15*\t};\foreach \x/\z in{-6*\s/3,-4*\s/2,-2*\s/1,0/3,2*\s/2,4*\s/1,6*\s/3}{
\draw (\x+\s,\y+\t) -- (\x,\y+2*\t)-- (\x-\s,\y+\t)-- (\x-\s,\y-\t)-- (\x,\y-2*\t)-- (\x+\s,\y-\t)-- cycle;
\draw(\x,\y)node{\z};}
\def\y{-18*\t};\foreach \x/\z in{-7*\s/2,-5*\s/1,-3*\s/3,-1*\s/2,1*\s/1,3*\s/3,5*\s/2,7*\s/1}{
\draw (\x+\s,\y+\t) -- (\x,\y+2*\t)-- (\x-\s,\y+\t)-- (\x-\s,\y-\t)-- (\x,\y-2*\t)-- (\x+\s,\y-\t)-- cycle;
\draw(\x,\y)node{\z};}
\def\y{-21*\t};\foreach \x/\z in{-6*\s/3,-4*\s/2,-2*\s/1,0/3,2*\s/2,4*\s/1,6*\s/3}{
\draw (\x+\s,\y+\t) -- (\x,\y+2*\t)-- (\x-\s,\y+\t)-- (\x-\s,\y-\t)-- (\x,\y-2*\t)-- (\x+\s,\y-\t)-- cycle;
\draw(\x,\y)node{\z};}
\def\y{-24*\t};\foreach \x/\z in{-3*\s/3,-1*\s/2,1*\s/1,3*\s/3}{
\draw (\x+\s,\y+\t) -- (\x,\y+2*\t)-- (\x-\s,\y+\t)-- (\x-\s,\y-\t)-- (\x,\y-2*\t)-- (\x+\s,\y-\t)-- cycle;
\draw(\x,\y)node{\z};}
\def\y{-27*\t};\foreach \x/\z in{0/3}{
\draw (\x+\s,\y+\t) -- (\x,\y+2*\t)-- (\x-\s,\y+\t)-- (\x-\s,\y-\t)-- (\x,\y-2*\t)-- (\x+\s,\y-\t)-- cycle;
\draw(\x,\y)node{\z};}
\def\x{0}\foreach \y in {-\t,-7*\t,-13*\t,-19*\t,-25*\t}{
\draw[->,red,very thick](\x,\y)--(\x,\y-\t);
\draw[->,red,very thick](\x,\y)--(\x+\s*\t,\y+0.5*\t);
\draw[->,red,very thick](\x,\y)--(\x-\s*\t,\y+0.5*\t);}
\foreach \x in {-3*\s,3*\s}{\foreach \y in {-4*\t,-10*\t,-16*\t,-22*\t}{
\draw[->,red,very thick](\x,\y)--(\x,\y-\t);
\draw[->,red,very thick](\x,\y)--(\x+\s*\t,\y+0.5*\t);
\draw[->,red,very thick](\x,\y)--(\x-\s*\t,\y+0.5*\t);}}
\foreach \x in {-6*\s,6*\s}{\foreach \y in {-7*\t,-13*\t,-19*\t}{
\draw[->,red,very thick](\x,\y)--(\x,\y-\t);
\draw[->,red,very thick](\x,\y)--(\x+\s*\t,\y+0.5*\t);
\draw[->,red,very thick](\x,\y)--(\x-\s*\t,\y+0.5*\t);}}
\def\x{1*\s};\def\y{-12*\t},
\fill[opacity=0.4,green] (\x+\s,\y+\t) -- (\x,\y+2*\t)-- (\x-\s,\y+\t)-- (\x-\s,\y-\t)-- (\x,\y-2*\t)-- (\x+\s,\y-\t)-- cycle;
\foreach \y in {-6*\t,-18*\t}{
\fill[opacity=0.4,red] (\x+\s,\y+\t) -- (\x,\y+2*\t)-- (\x-\s,\y+\t)-- (\x-\s,\y-\t)-- (\x,\y-2*\t)-- (\x+\s,\y-\t)-- cycle;}
\foreach \x in {-2*\s,4*\s}{\foreach \y in {-9*\t,-15*\t}{
\fill[opacity=0.4,red] (\x+\s,\y+\t) -- (\x,\y+2*\t)-- (\x-\s,\y+\t)-- (\x-\s,\y-\t)-- (\x,\y-2*\t)-- (\x+\s,\y-\t)-- cycle;}}
\end{tikzpicture}}
\caption{Frequency reuse pattern}\label{fig:freqreuse}
\end{center}
\end{figure}

The frequency reuse index is three, that is, at each cluster of three cells, 
its antenna sector uses one of the three available frequencies. 
In our propagation model we have only considered the interference from the six closest cells, 
assuming that the interference from more distant cells is negligible. 
Fig.~\ref{fig:freqreuse} depicts the frequency reuse pattern, 
showing that every cell interferes with one cell at each cluster surrounding its own cluster. 
In this figure, the colored cells use frequency 1, 
and the green cell in the center receives interference power from the six red cells. 

The capacity of cell $j$ perceived by a user at a given position in cell $j$ is considered to be equal to the Shannon capacity, 
\begin{equation}
c^{(j)}= \text{BW}  \log_2\left( 1 + \text{SINR}\right) \; \text{bps},
\end{equation}
where $\text{BW}$ is the transmission bandwidth
and SINR is the signal-to-interference-plus-noise ratio 
\begin{equation}
\text{SINR}=\frac{P_r}{N+I_r},
\end{equation}
where $P_r$ is the received power,
$N$ is the thermal noise, 
and $I_r$ is the received interference.

SINR is computed based on the physical layer network model specified in \cite{itu2009}.
The received power is calculated from the transmitted power, the antenna gain, the path loss
and a log-normal shadow fading with zero mean and standard deviation of 4 dB.
The antenna gain depends on the angle to the antenna~\eqref{eq:antennaGain} and 
the path loss depends on the distance to the antenna, 
with a minimum distance of $10$\,m and assuming a carrier frequency of 2.5 GHz.
The thermal noise is $-104$\,dBm, for $\text{BW}=10$\,MHz and a power density of $-174$\,dBm/Hz.
The interference power is calculated as the sum of the signal powers 
received from the six cells using the same frequency at the six surrounding cell clusters. 

Users move according to the Random Waypoint Model (RWP),
in which they alternate between a \texttt{paused} state and a \texttt{moving} state. 
Users remain in the \texttt{paused} state a random time uniformly distributed in $[0,t_{\text{Pmax}}]$,
where $t_{\text{Pmax}}$ is the maximum pause time.
At \texttt{moving} state, users travel in a straight line at a constant speed of 3 km/h
with a random direction uniformly distributed in $[0,2\pi]$.
Every time a user enters the \texttt{moving} state, 
it travels during a random timed uniformly distributed in $[0,t_{\text{Wmax}}]$,
where $t_{\text{Wmax}}$ is the maximum walk time.

A handover occurs when a user crosses the cell boundary. According to the wrap-around
configuration, 
if a user exits the boundary of the 57 cells grid, 
it enters the cell at the opposite side of the grid.

At the start of the simulation, 
a certain number of users are located at each cell
at random locations uniformly distributed over the cell. 
All the users remain active for the entire simulation, 
moving through the cell grid according to the WRP pattern.

\subsection{Subscription model}\label{sec:usermodel}

At the start of the simulation, 
every user subscribes to an NST and hereafter the subscriptions are
periodically updated every subscription time $t_s$,
and every time a handover occurs.
Every time that a user subscribes or updates the subscription, the chosen NST is the one that provides the greatest utility to the user, 
considering for this the sum of 
the observed part of the utility given by~\eqref{eq:utility} plus the unobserved user-specific part of the utility.

The unobserved user-specific part of the utility $\kappa_{u,i}$ is obtained for each user $u$ and each NST $i$ 
at the start of the simulation from a Gumbel distribution of zero mean and parameter $\nu$. 
This value models the  perception that each user has of the unobserved part of the utility provided by each NST, 
and depends only on the user-NST pair. 
We assume that, for each user-NST pair, $\kappa_{u,i}$ is the same at all the cells and does not change during the simulation.
The observed part of the utility is calculated for all NSTs every time that a user updates the subscription.
This part of the utility depends on $r^{(j)}_i$, 
which is the bit rate that NST~$i$ assigns to users in cell~$j$ (the current cell of the user updating the subscription), 
according to
\begin{equation}\label{eq:utilij}
r_i^{(j)}=
\frac{\omega^{(j)}_i}{\sum_{t \in \mathcal{S}} \omega^{(j)}_t} \frac{c_u^{(j)}}{n_i^{(j)}} \qquad i \in \mathcal{S},\qquad j \in \mathcal{\textbf{\textbf{B}}},
\end{equation}
where $c_u^{(j)}$ is the cell capacity at the user’s current position. 
Note that the factor $\omega^{(j)}_i/(\sum_{t \in \mathcal{S}}\omega_i^{(j)}) n^{(j)}_t$ in~\eqref{eq:utilij} depends only on the weights assigned by each NST to cell~$j$ 
and the number of subscribers to each NST at cell~$j$.
This information can be provided to the user by the NSTs.
In contrast, the data transmission capacity at cell~$j$, $c_u^{(j)}$, depends on the user's position, 
so it is different for each user.
Therefore, users at cell~$j$ will have to separately estimate a value of $c^{(j)}$
by keeping track of the capacity that they measure throughout their trajectory through the cell~$j$. 
Let $\hat{c}_u^{(j)}$ be the cell capacity estimated by user~$u$ at cell~$j$.
User~$u$ updates $c_u^{(j)}$ every update time $t_{\text{update}}$ by means of
\begin{equation}\label{eq:capFilter}
\hat{c}_u^{(j)}\leftarrow (1- \lambda) c_u^{(j)}+ \lambda c_{u,\text{last}}^{(j)},
\end{equation}  
where $c_{u,\text{last}}^{(j)}$ is the last measure of cell~$j$'s capacity
made by user~$u$, 
and $\lambda<1$ is a memory parameter 
which defines an exponential moving average (EMA) 
of the successive values of the capacity measured by the user.
The $\lambda$ parameter models the amount of memory 
that a user keeps about the capacity perceived at the current cell;
a low value of $\lambda$ results in a long-term estimation of the capacity 
that gets closer to the spatial average of the cell capacity, 
while a high  value of $\lambda$ results in a fast changing value 
of the capacity.

Every time that user~$u$ moves a distance $d_{\text{update}}$ 
from the position in which the last measure of  $c_{u,\text{last}}^{(j)}$ was made, 
a new measure of  $c_{u,\text{last}}^{(j)}$ is taken.
The update distance $d_{\text{update}}$ must be such that the received SINR 
has changed significantly compared to the previous measure 
and the shadow fading is uncorrelated with the one of the previous measure. 

\begin{figure}[htb]
\begin{center}
\resizebox{0.8\columnwidth}{!}{
\begin{tikzpicture}[scale=0.8, transform shape]
	\tikzset{node style/.style={state, 
	minimum width=0.3in,
	line width=0.2mm,
	text width=4em, 
	align=center,	}}	
	\node[] at (0, 0) (init) {};	
	\node[node style] at (0, -4) (paused) {\texttt{paused}};	
	\node[node style] at (0, -7) (moving) {\texttt{moving}};	
	\draw[every loop,	
	    line width=0.2mm,
	>=latex,		]	    	    
	 (init) edge[] node[text width=3.5cm,align=center,right] {user starts\break [drawss $\kappa_{u,i}$,\break measures $c_{u,\text{last}}^{(j)}$,\break $\hat{c}_u^{(j)}=c_{u,\text{last}}^{(j)}$,\break subscribes]} (paused)
	 (paused) edge[bend right=40] node[text width=3.5cm,align=center,left] {end of pause time} (moving)
	 (moving) edge[bend right=40] node[text width=3.5cm,align=center,right] {end of walk time}  (paused)	
	 (paused) edge[text width=3.7cm,align=center,loop left] node {end of subscription time\break [subscribes]} (paused)	
	 (paused) edge[text width=3cm,align=center,loop right] node {end of update time\break [updates $\hat{c}_u^{(j)}$]} (paused)	
	 (moving) edge[text width=3.7cm,align=center,loop left] node {end of subscription time\break [subscribes]} (moving)	
	 (moving)  edge[text width=3cm,align=center,loop right] node {end of update time\break [updates $\hat{c}_u^{(j)}$]} (moving)
	  (moving)  edge[text width=4.5cm,align=center,left,in=-140,out=-105,looseness=6 ] node {20 m away from last measure\break [measures $c_{u,\text{last}}^{(j)}$]} (moving)	
	   (moving)  edge[text width=2.2cm,align=center,right,in=-75,out=-40,looseness=6 ] node {handover\break [subscribes]} (moving)		;	    
	\end{tikzpicture}
}
	\caption{User state diagram}\label{fig:userstates}
\end{center}
\end{figure}
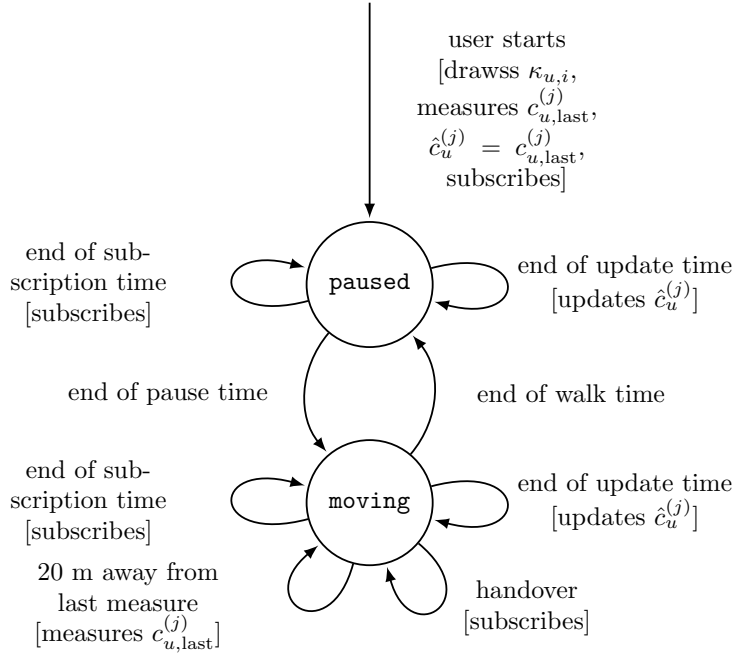

The state diagram of Fig.~\ref{fig:userstates} summarizes the user behavior.

\begin{table}[tb]
\begin{center}
\caption{Simulation parameters}\label{table:parameters}
\renewcommand{\arraystretch}{1.1}
\resizebox{0.8\columnwidth}{!}{
\begin{tabular}{ll}
\toprule 
Parameters fixed for all simulations & Value \\ \hline
Number of cells, $|\mathcal{B}|$ & 57 \\
Inter station distance, $\text{ISD}$ & 200 m  \\
Power transmitted, $P_0$ & 41 dBm \\
Maximum antenna gain, $G_{\text{max}}$ & 17 dB \\
3 dB beamwidth, $\theta_{3\text{dB}}$ & $70\pi/180$ rads \\
Maximum antenna attenuation, $A_m$ & 20 dB \\
Transmission bandwidth, $\text{BW}$ & 10 MHz \\
Carrier frequency & 2 GHz \\
Thermal noise power density & -174 dB/Hz \\
Log-normal shadow fading deviation & 4 dB\\
User velocity in \texttt{moving} state & 3 km/h \\
Utility parameter, $\mu$ & 2\\
Gumbel parameter $\nu$ & 1\\
Maximum pause time $t_{\text{Pmax}}$ & 120 s  \\
Maximum walk time $t_{\text{Wmax}}$ & 120 s \\
Distance to update $c_{u, \text{last}}^{(j)}$, $d_{\text{update}}$ & 20 m \\
Time between $\hat{c}^{(j)}$ updates, $t_{\text{update}}$ & 24 s \\  \\
Parameters varied & Reference value \\ \hline
Time average of the number of users per cell, $\bar{n}$& 250 \\
Number of NTSs, $|\mathcal{S}|$ & 4 \\
NSTs shares, $\{s_i\}$ &$\{0.1,0.2,0.3,0.4\}$ \\
Weights set by NSTs, $\omega_i$ & $\{\frac{0.1}{57},\frac{0.2}{57},\frac{0.3}{57},\frac{0.4}{57}\}$\\
No-subscription bit rate, $r_0$ & 500 kpbs \\
EMA parameter $\lambda$ & 0.1\\
Subscription time, $t_s$ &  240 s \\  
\bottomrule
\end{tabular}}
\end{center}
\end{table}

\section{Results and discussion}\label{sec:results}

In this section the subscription ratios and the fractions of subscribers to each NST 
obtained by simulation (as specified in Section~\ref{sec:problemSim} and Section~\ref{sec:simmodel}) are compared 
with those obtained analytically from~\eqref{eq:penetration_NLeq} and~\eqref{eq:n_rs}. 

The simulation results have been obtained for a setup in which  
the utility parameter is $\mu=2$,
the Gumbel distribution parameter is $\nu=1$,
the maximum walk time has been set to $t_{\text{Wmax}}=120$ s, 
corresponding to a distance of $100$~m at a speed of $3$ km/h,
the maximum pause time has been set to $t_{\text{Pmax}}=t_{\text{Wmax}}$,
the user measures the cell capacity when it moves a distance $d_{\text{update}}$ 
($20$ m) away from the point where the last measure was taken,
and the time to update the cell capacity is set to $t_{\text{update}}=24$ s, 
which is the time corresponding to a traveling distance $d_{\text{update}}$  at a user speed of $3$ km/h.

A reference network configuration have been set up in which
the number of NSTs is $|\mathcal{S}|=4$ with shares $\{s_i\}=\{0.1,0.2,0.3,0.4\}$, 
the weights set by the NSTs at every cell are $\omega_i^{(j)}=\omega_i=s_i/|\mathcal{B}|$,
the no-subscription bit rate is  $r_0 = 500$\,kbps, 
the EMA parameter is $\lambda=0.1$,
the subscription time is related to $\lambda$ and $t_{\text{update}}$ through $\lambda t_s=t_{\text{update}}$.
All cells start the simulation with $250$ users each, 
and due to the spatial symmetry of the grid and the stationarity of the mobility parameters, 
the time average of users in cell~$j$,  $\bar{n}^{(j)}$, converges to $250$ users for all the cells ($\bar{n}^{(j)}=\bar{n}=250)$.

The simulation parameters and the reference network configuration are summarized in Table~\ref{table:parameters}.

Since in this configuration all cells have the same parameters, 
the results will also be identical in all of them. 
Then, from now on we will denote by $\bar{c}$ and $\hat{c}$ the estimates, 
of the mean and the median, respectively, of the capacity at any cell, 
by $\tilde{\beta}$ the sensitivity parameter of the modified model at any cell,
by $\sigma(\bar{c})$, $\sigma(\hat{c})$ and $\sigma(\tilde{\beta})$ the subscription ratios at any cell, 
and by  $\rho_i$ and $\rho_i(\tilde{\beta})$ the fractions of subscribers to NST~$i$ at any cell.

Starting from the reference configuration described above, several experiments have been conducted, 
in order to determine the influence of different parameters in the results. 
Specifically, simulations for five cases have been carried out,
varying a parameter in each one of them 
while keeping the rest of the parameters as in the reference configuration. 
These five cases are:
\begin{enumerate}[label=\alph*)]
\item The time average of the number of users per cell, $\bar{n}$, is varied in the set \\$\{100, 150, 200, 250, 300, 350\}$.
\item The number of NSTs, $|\mathcal{S}|$, is varied in the set $\{2, 3, 4, 5, 6, 7\}$. 
For each value of $|\mathcal{S}|$, the shares of the NSTs are set to $\{s_i\}=\{1/k\dots |\mathcal{S}|/k\}$ 
with $k=|\mathcal{S}|(|\mathcal{S}|+1)/2$, so that $\sum_{i \in \mathcal{S}}s_i=1$.
\item The parameter that defines the no-subscription option, $r_0$ (in kbps), is varied in the set $\{200, 300, 400, 500, 600, 700\}$.
\item The EMA parameter $\lambda$, is varied in the set $\{0.1, 0.15, 0.2, 0.25, 0.3, 0.35\}$. 
Since in the reference configuration $t_s$ was set in relation to $\lambda$, 
in this case the product $\lambda t_s$ is set to its reference value $\lambda t_s=24$ s for each value of $\lambda$.b
\item The product of the subscription time and the EMA parameter, $\lambda t_s$ (in seconds), is varied in the set $\{12, 24, 36, 48, 56, 72\}$.
\end{enumerate}

In order to compare the simulation results with the analytic ones, 
for each case, three sets of analytic values have been obtained:
\begin{enumerate}
\item The subscription ratios $\sigma(\bar{c})$ 
obtained by using in~\eqref{eq:penetration_NLeq} an estimate of the mean capacity;
and the fractions of subscribers to each NST, $\rho_i$, obtained from~\eqref{eq:n_rs}.
\item The subscription ratios $\sigma(\hat{c})$  obtained using in~\eqref{eq:penetration_NLeq} an estimate of the median of
the capacity. The fractions of subscribers in this result set are identical to those in
the first result set ($\rho_i$), since~\eqref{eq:n_rs} does not depend on the cell capacity.
\item The subscription ratios $\sigma(\tilde{\beta})$ 
and the fractions of subscribers  $\rho_i(\tilde{\beta})$
obtained as in the first set of results, 
using $\bar{c}$ in~\eqref{eq:penetration_NLeq},
but now taking the users sensitivity parameter $\tilde{\beta}$ from the modified model~\eqref{eq:alfaModified}.
The estimate of the variance of the capacity, $\Var[\log(c)]$, 
used in~\eqref{eq:alfaModified}, 
is obtained in the same way as  $\bar{c}$, from  the cell capacities perceived by the users at the subscription times.
\end{enumerate}

\begin{figure}[tb]
\begin{center}
\includegraphics[width=0.6\columnwidth]{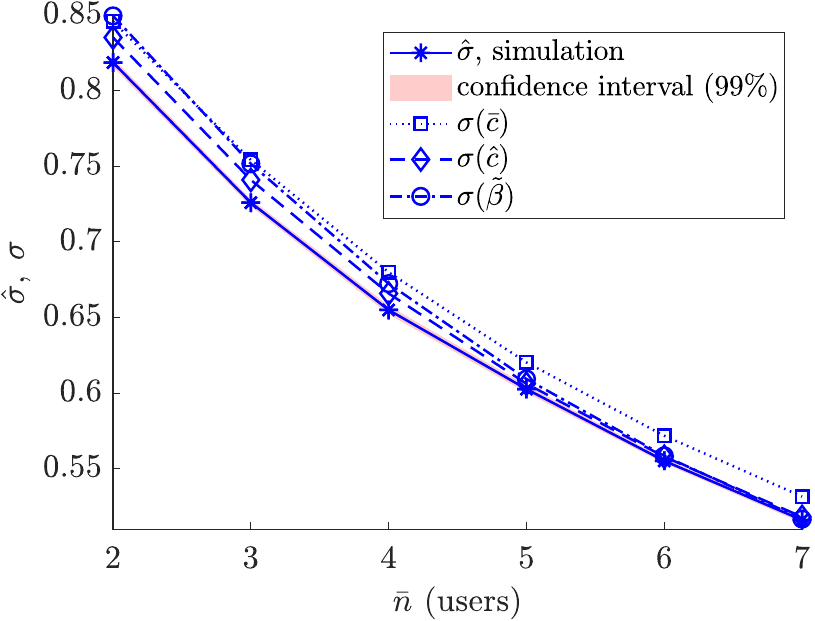}
\caption{Subscription ratios, simulated ($\hat{\sigma}$) and analytic ($\sigma$), 
varying the time average of the number of users per cell $\bar{n}$ (case a).}
\label{fig:sigmausers}
\end{center}
\end{figure}

\begin{figure}[tb]
\begin{center}
\includegraphics[width=0.6\columnwidth]{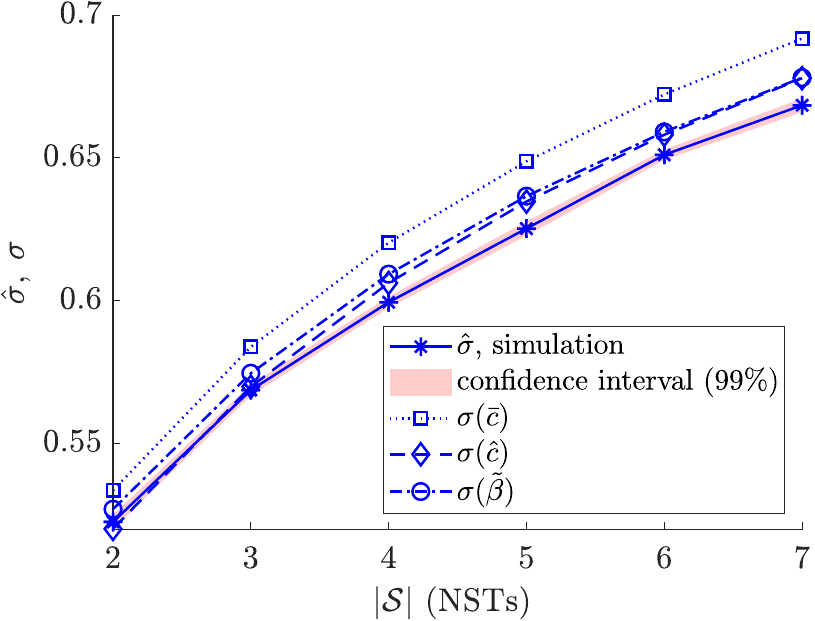}
\caption{Subscription ratios, simulated ($\hat{\sigma}$) and analytic ($\sigma$), 
varying the number of NSTs $|\mathcal{S}|$ (case b).}
\label{fig:sigmaslices}
\end{center}
\end{figure}

\begin{figure}[tb]
\begin{center}
\includegraphics[width=0.6\columnwidth]{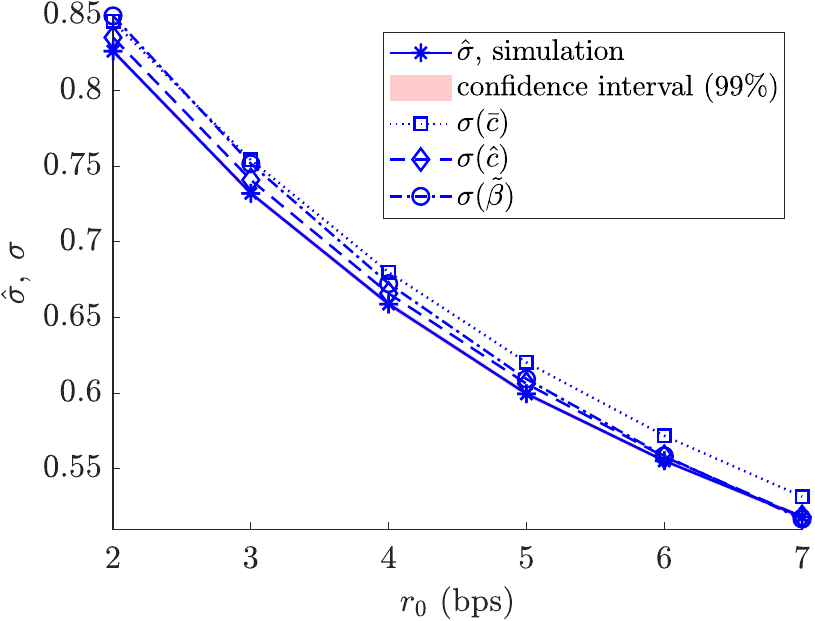}
\caption{Subscription ratios, simulated ($\hat{\sigma}$) and analytic ($\sigma$), 
varying the  no-subscription option parameter $r_0$ (case c).}
\label{fig:sigmar0}
\end{center}
\end{figure}

\begin{figure}[tb]
\begin{center}
\includegraphics[width=0.6\columnwidth]{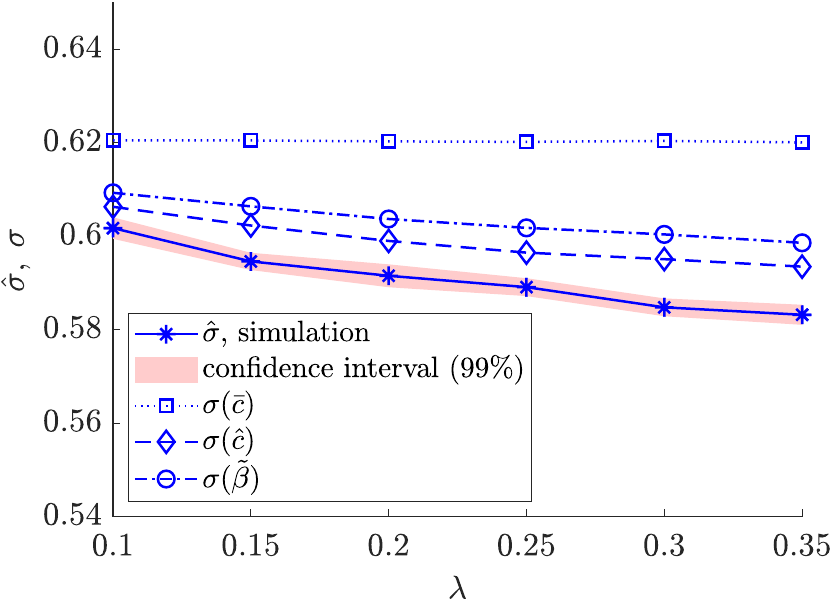}
\caption{Subscription ratios, simulated ($\hat{\sigma}$) and analytic ($\sigma$), 
varying the EMA parameter $\lambda$ (case d).}
\label{fig:sigmaalpha}
\end{center}
\end{figure}

\begin{figure}[tb]
\begin{center}
\includegraphics[width=0.6\columnwidth]{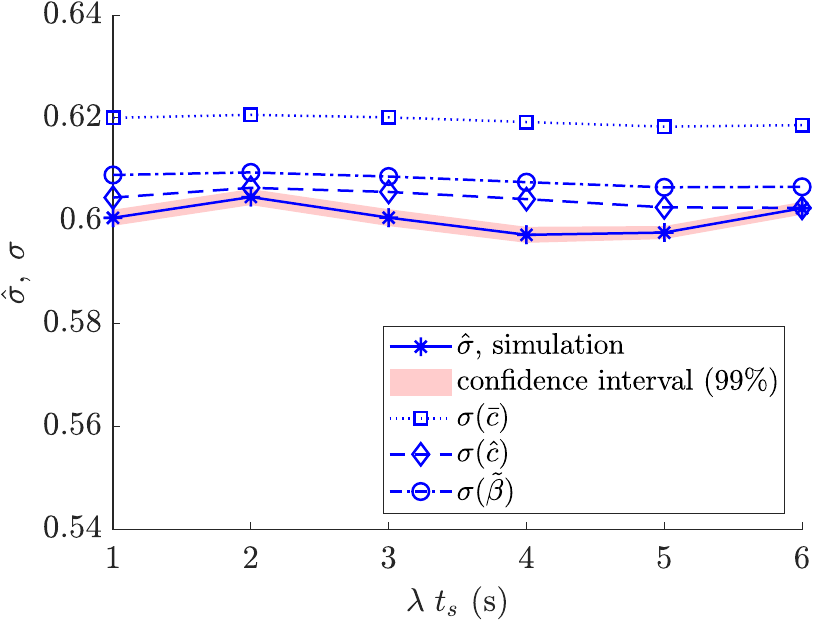}
\caption{Subscription ratios, simulated ($\hat{\sigma}$) and analytic ($\sigma$), 
varying the the subscription time $t_S$ (case e).}
\label{fig:sigmasusc}
\end{center}
\end{figure}

Fig.~\ref{fig:sigmausers}, Fig.~\ref{fig:sigmaslices}, Fig.~\ref{fig:sigmar0}, Fig.~\ref{fig:sigmaalpha} and Fig.~\ref{fig:sigmasusc} 
show the subscription ratios,
both simulated and analytic, obtained in the five cases.
For the results obtained from simulation, the interval for a confidence level of $99\%$ is also plotted.

\begin{figure}[tb]
\begin{center}
\includegraphics[width=0.6\columnwidth]{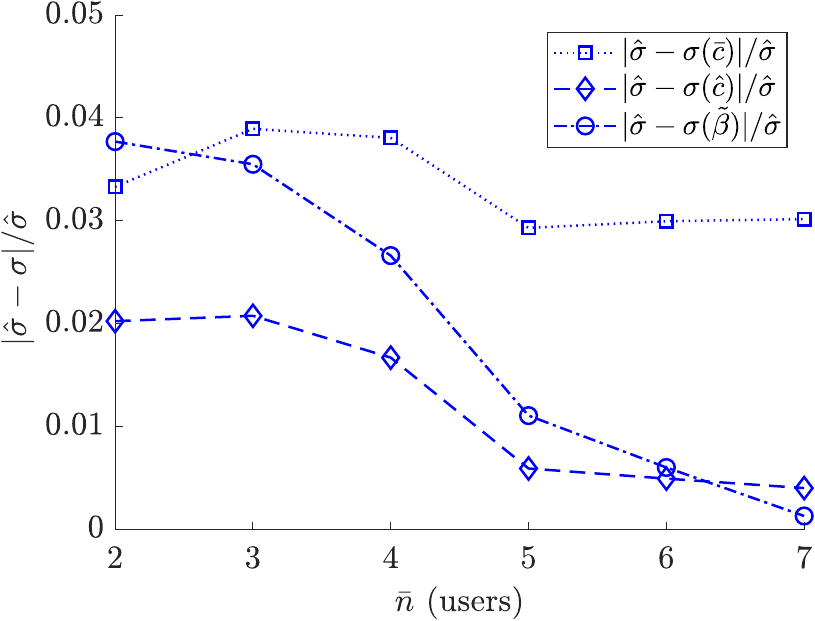}
\caption{Relative error of the subscription ratio for each set of analytic results, 
varying the time average of the number of users per cell $\bar{n}$ (case a).}
\label{fig:sigmaErrorusers}
\end{center}
\end{figure}

\begin{figure}[tb]
\begin{center}
\includegraphics[width=0.6\columnwidth]{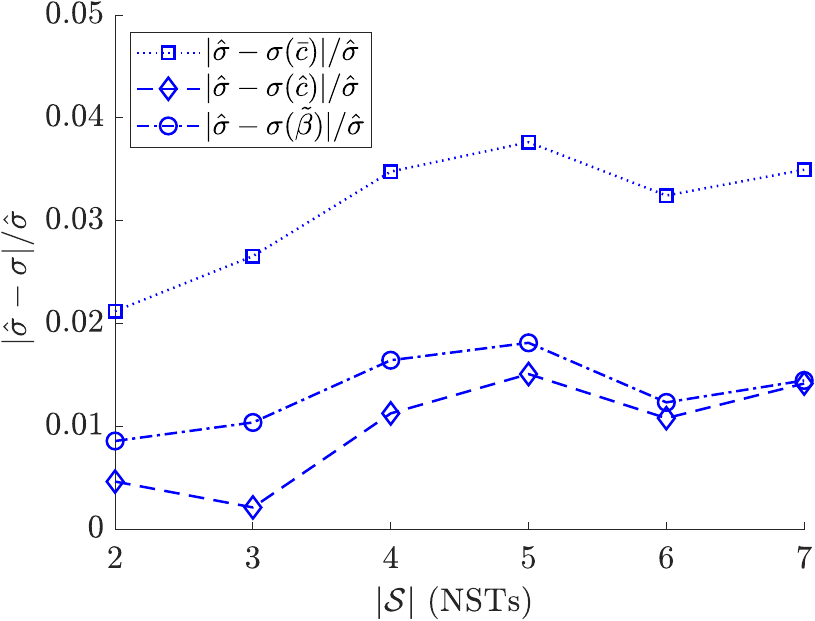}
\caption{Relative error of the subscription ratio for each set of analytic results, 
varying the number of NSTs $|\mathcal{S}|$ (case b).}
\label{fig:sigmaErrorslices}
\end{center}
\end{figure}

\begin{figure}[tb]
\begin{center}
\includegraphics[width=0.6\columnwidth]{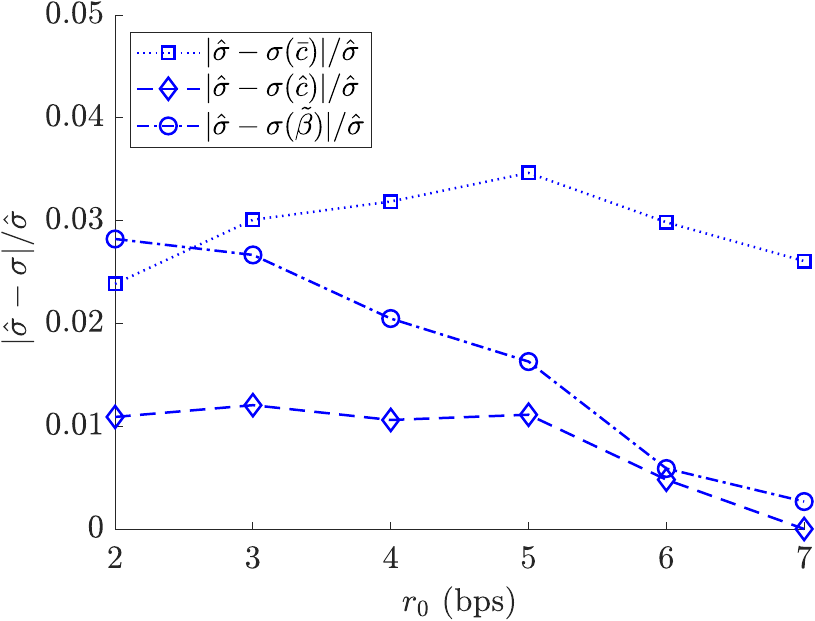}
\caption{Relative error of the subscription ratio for each set of analytic results, 
varying the  no-subscription option parameter $r_0$ (case c).}
\label{fig:sigmaErrorr0}
\end{center}
\end{figure}

\begin{figure}[tb]
\begin{center}
\includegraphics[width=0.6\columnwidth]{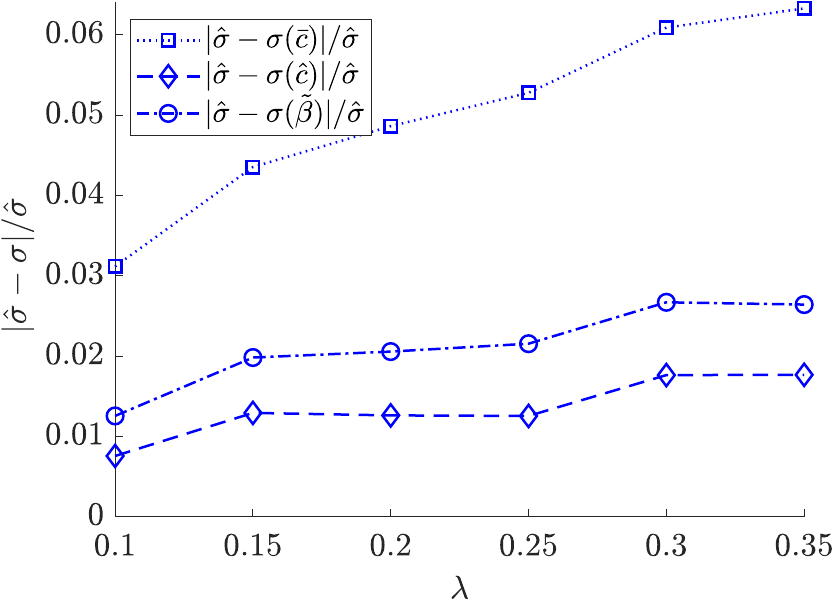}
\caption{Relative error of the subscription ratio for each set of analytic results, 
varying the EMA parameter $\lambda$ (case d).}
\label{fig:sigmaErroralpha}
\end{center}
\end{figure}

\begin{figure}[tb]
\begin{center}
\includegraphics[width=0.6\columnwidth]{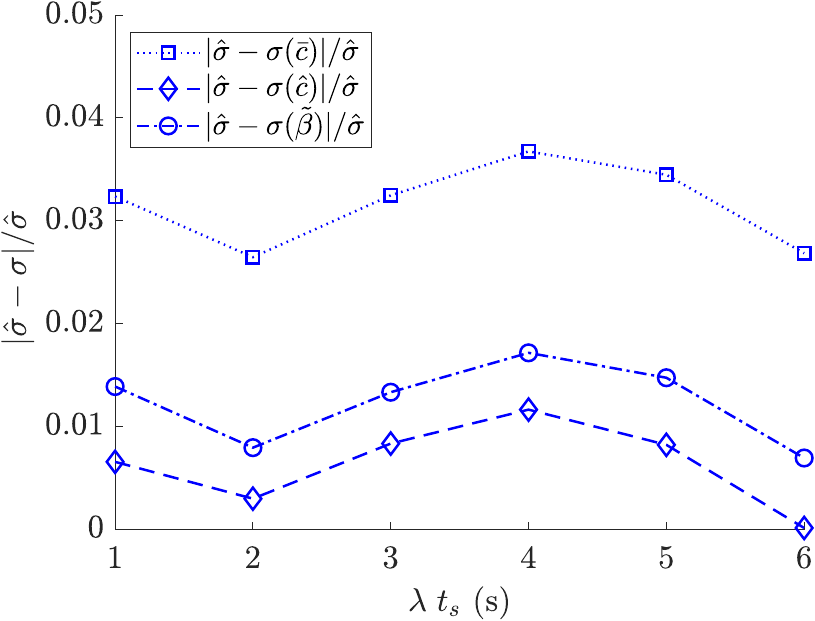}
\caption{Relative error of the subscription ratio for each set of analytic results, 
varying the the subscription time $t_S$ (case e).}
\label{fig:sigmaErrorsuscr}
\end{center}
\end{figure}

The accuracy of each set of analytic results is better appreciated in Fig.~\ref{fig:sigmaErrorusers},
Fig.~\ref{fig:sigmaErrorslices}, Fig.~\ref{fig:sigmaErrorr0}, Fig.~\ref{fig:sigmaErroralpha} and Fig.~\ref{fig:sigmaErrorsuscr}, 
which show the error of the subscription ratios obtained analytically
relative to the value obtained by simulation, computed as
\begin{equation}
\text{relative error of}\; \sigma \triangleq \frac{|\hat{\sigma}-\sigma|}{\hat{\sigma}},
\end{equation}
where $\sigma=\sigma(\bar{c})$, or $\sigma=\sigma(\hat{c})$, or $\sigma=\sigma(\tilde{\beta})$.

The subscription ratios obtained with the mean of the capacity, $\sigma(\bar{c})$,
are the ones with the largest error.
Therefore, it does not seem that the mean of the capacity 
is the best statistic to approximate the cell capacity. 
The results obtained with the median, $\sigma(\hat{c})$, 
and the ones obtained with the modified model, $\sigma(\tilde{\beta})$, 
are very close to each other, and clearly outperform those obtained with the mean. 
In most cases, both of them show a relative error below $2\%$,
and  $\sigma(\hat{c})$ slightly outperforms  $\sigma(\tilde{\beta})$.

For each of the cases described above, 
what these results indicate about the influence of the parameter under study 
on the accuracy of the model is the following:
\begin{enumerate}[label=\alph*)]
\item A larger number of users per cell results in a greater accuracy of the model. 
For cells with more than 300 users the error is negligible. 
A high number of users in the cell seems to alleviate the effect of the variance in the capacity. 
\item The number of NSTs does not affect significantly the accuracy of the results.
\item The greater the value of $r_0$, the greater the accuracy of the model. 
The model works better when the fraction of non-subscribing users is not too small.
\item The lower the value of $\lambda$, the greater the accuracy of the model. 
A lower $\lambda$ means more memory in the user estimate of the cell capacity,  
which results in a smaller variation in the capacity used for the subscription.
\item 
For some ranges of $\lambda t_s$ the error increases and for others it decreases. Therefore,
from these results, we cannot draw general conclusions on the effect of the subscription time on the error.
\end{enumerate} 

\begin{table}[htb]
\centering
\caption{NST shares and fractions of subscribers (analytic results)}
\resizebox{0.6\columnwidth}{!}{
\begin{tabular}{lcccccc}
\toprule 
\multicolumn{7}{c}{\vspace{-4mm}} \\
 &  $|\mathcal{S}|=2$ & $|\mathcal{S}|=3$ & $|\mathcal{S}|=4$  & $|\mathcal{S}|=5$ &  $|\mathcal{S}|=6$  & $|\mathcal{S}|=7$   \\
\multicolumn{7}{c}{\vspace{-2mm}} \\ \hline
$s_1$ & 0.333 & 0.167 & 0.100 & 0.067 & 0.048 & 0.036\\
$s_2$ & 0.667 & 0.333 &  0.200 & 0.133 &   0.095 &  0.071 \\ 
$s_3$ &  &  0.500 & 0.300 & 0.200 & 0.143 & 0.107 \\ 
$s_4$ &  &  & 0.400 & 0.267 & 0.190 &  0.143 \\
$s_5$ &  &  &  & 0.333 & 0.238 & 0.179  \\
$s_6$ &  &  &  &  & 0.286 & 0.214  \\
$s_7$ &  &  &  &  &  &  0.250\\ \hline
$\rho_1$ & 0.387 & 0.214 & 0.139 & 0.099 & 0.075 & 0.059  \\ 
$\rho_2$ & 0.613 & 0.340 &  0.221 & 0.157  & 0.118 &  0.093\\
$\rho_3$ &  &  0.446  &  0.289& 0.206 &  0.155& 0.122 \\
$\rho_4$ &  &  &  0.351  &  0.249 &  0.188& 0.148 \\
$\rho_5$ &  &  &  &   0.289 & 0.218 & 0.171 \\
$\rho_6$ &  &  &  &  &   0.246 & 0.193 \\
$\rho_7$ &  &  &  &  &  &   0.214 \\ \hline
$\rho_1(\tilde{\beta})$ &  0.383 &  0.211 &  0.136 & 0.097 & 0.073 & 0.057  \\
$\rho_2(\tilde{\beta})$ &  0.617 & 0.340 &  0.220 & 0.155 & 0.117 & 0.091  \\
$\rho_3(\tilde{\beta})$ &  &  0.449  &   0.290 & 0.205  & 0.154 &  0.121\\
$\rho_4(\tilde{\beta})$ &  &  & 0.354 & 0.251 &  0.188 & 0.147 \\
$\rho_5(\tilde{\beta})$ &  &  &  & 0.292 &  0.219 &  0.172 \\
$\rho_6(\tilde{\beta})$ &  &  &  &  & 0.249 &  0.195 \\
$\rho_7(\tilde{\beta})$ &  &  &  &  &  &   0.217\\
\bottomrule
\end{tabular}}\label{table:shares}
\end{table}

\begin{table}[htb]
\centering
\caption{Relative error of the fractions of subscribers  (average of all slices)}
\resizebox{0.6\columnwidth}{!}{
\begin{tabular}{llllllcc}
\toprule 
\multicolumn{8}{c}{\vspace{-4mm}} \\
case & $\bar{n}$ & $|\mathcal{S}|$ &$r_0$ & $\lambda$ & $\lambda t_s$&  $\frac{\vert\hat{\rho}_i-\rho_i \vert}{\hat{\rho}_i}$ & 
$\frac{\vert\hat{\rho}_i-\rho_rhoi(\tilde{\beta}) \vert}{\hat{\rho}_i}$  \\
\multicolumn{8}{c}{\vspace{-2mm}} \\ \hline
\multirow{6}{*}{a)} & $100$ & \multirow{6}{*}{$4$}& \multirow{6}{*}{$500$}  & \multirow{6}{*}{$0.1$} &  \multirow{6}{*}{$24$} 
      & $ 0.005$ & $ 0.013$  \\ 
& $150$ & & & & & $ 0.009$ & $ 0.006$ \\ 
& $200$ & & & & & $ 0.007$ & $ 0.006$  \\ 
& $250$ & & & & & $ 0.004$ & $ 0.009$ \\ 
& $300$ & & & & & $ 0.007$ & $ 0.009$  \\ 
& $350$ & & & & & $ 0.013$ & $ 0.007$ \\  \cmidrule{1-6}     
\multirow{6}{*}{b)} &\multirow{6}{*}{$250$} &$2$ & \multirow{6}{*}{$500$} & \multirow{6}{*}{$0.1$} & \multirow{6}{*}{$24$}  
             & $ 0.002$ & $ 0.006$  \\ 
& & $3$  & & & & $ 0.004$ & $ 0.005$ \\ 
& & $4$  & & & & $ 0.007$ & $ 0.005$ \\ 
& & $5$  & & & & $ 0.003$ & $ 0.011$ \\ 
& & $6$  & & & & $ 0.011$ & $ 0.013$  \\ 
& & $7$  & & & & $ 0.010$ & $ 0.006$ \\   \cmidrule{1-6}    
\multirow{6}{*}{c)} &\multirow{6}{*}{$250$} & \multirow{6}{*}{$4$} & $200$ & \multirow{6}{*}{$0.1$} & \multirow{6}{*}{$24$}  
              & $ 0.007$ & $ 0.009$  \\ 
& & & $300$ & & & $ 0.004$ & $ 0.008$  \\ 
& & & $400$ & & & $ 0.004$ & $ 0.006$ \\ 
& & & $500$ & & & $ 0.003$ & $ 0.007$  \\ 
& & & $600$ & & & $ 0.005$ & $ 0.008$  \\ 
& & & $700$ & & & $ 0.008$ & $ 0.008$  \\   \cmidrule{1-6}   
\multirow{6}{*}{d)} &\multirow{6}{*}{$250$} & \multirow{6}{*}{$4$} & \multirow{6}{*}{$500$} & $0.1$ & \multirow{6}{*}{$24$}  
              & $ 0.002$ & $ 0.010$  \\ 
& & & & $0.2$ & & $ 0.011$ & $ 0.010$  \\ 
& & & & $0.2$ & & $ 0.008$ & $ 0.013$ \\ 
& & & & $0.2$ & & $ 0.008$ & $ 0.017$  \\ 
& & & & $0.3$ & & $ 0.005$ & $ 0.012$  \\ 
& & & & $0.4$ & & $ 0.008$ & $ 0.015$ \\  \cmidrule{1-6}   
\multirow{6}{*}{e)} &\multirow{6}{*}{$250$} & \multirow{6}{*}{$4$} & \multirow{6}{*}{$500$} & \multirow{6}{*}{$0.1$} & $12$
             & $ 0.008$ & $ 0.008$ \\ 
& & & & & $24$ & $ 0.005$ & $ 0.009$  \\ 
& & & & & $36$ & $ 0.007$ & $ 0.008$  \\ 
& & & & & $48$ & $ 0.005$ & $ 0.011$ \\ 
& & & & & $60$ & $ 0.005$ & $ 0.012$  \\ 
& & & & & $72$ & $ 0.005$ & $ 0.013$  \\  
\bottomrule
\end{tabular}}\label{table:rho}
\end{table}

The analytic values of the fractions of subscribers for each NST, ${\rho_i}$ and ${\rho_i(\tilde{\beta})}$,
are shown in Table~\ref{table:shares}. Note that these are the same in all the cases studied, except in
case b), since they only depend on the weights of the NSTs~\eqref{eq:n_rs}, which have been chosen
depending in $s_i$ only.

Table~\ref{table:rho}, shows the average relative error of each set of analytic results
in the five cases. 
Here it can be seen that, notwithstanding the variation in the utility, 
the model provides an accurate estimation of the fractions of subscribers.
Most values on the table exhibit a relative error for $\rho$
below $0.1\%$, clearly below the relative error of $\sigma$, 
and the accuracy achieved by $\rho_i(\tilde{\beta})$ (with the modified model) is very similar~\eqref{eq:n_rs} to that obtained with the unmodified model.

\section{Conclusions}\label{sec:conclusion}

The logit model has been evaluated in a business model for network slicing in mobile radio access networks, 
under the assumption that 
in a given cell all subscribers to a given NST obtain the same bit rate.
To calculate this bit rate, the number of subscribers to an NST in a cell was assumed to be constant 
and equal to a time average of the number of subscribers, 
and the capacity was also assumed to be constant and equal to an average of the actual cell capacity seen by all users in this cell.

Three alternatives to implement this assumption have been tested: 
a first in which the model uses the mean of the cell capacity; 
a second in which the model uses the median of the cell capacity; 
and a third in which the mean of the cell capacity is used and also 
the logit model has been modified in order to include the effect of the variation of the cell capacity 
in the variation of the unobserved part of the utility.

The analytical results obtained with this model have been compared 
with the results obtained by simulating a realistic setup 
which includes user mobility and a characterization of radio propagation.
This comparison shows that the logit model is robust enough 
to provide valid results for scenarios of mobile radio communications.

From the three alternatives tested, 
the one in which the model uses the median of the cell capacity
is the one that more accurately models the business in a radio access mobile environment. 
If an estimate of the median of the cell capacity is used to approximate the observed part of the user utility, 
the logit model provides very accurate results for the fractions of subscribers to each NST (relative error below $0.1\%$), 
and quite accurate results (relative error below $2\%$) for the subscribing ratios.

These results generalize the validity of the results  in~\cite{guijarro2021} 
to scenarios with mobility and a highly varying radio channel capacity because of radio propagation. 
This generalization is particularly valid and provide very accurate results 
in those cases where at least one of the following conditions holds:
\begin{itemize}
\item There is a large number of users per cell.
\item The fraction of unsubscribed users is not very low.
\item The users estimate the cell capacity with a long memory EMA parameter.
\end{itemize}

\section*{Funding sources}

This work was supported by the Spanish Ministry of Science and Innovation\\ (MCIN/AEI/10.13039/501100011033), 
and the European Union (ERDF, A way of making Europe) through Grant PGC2018-094151-B-I00.


\begin{thebibliography}{10}
\expandafter\ifx\csname url\endcsname\relax
  \def\url#1{\texttt{#1}}\fi
\expandafter\ifx\csname urlprefix\endcsname\relax\def\urlprefix{URL }\fi
\expandafter\ifx\csname href\endcsname\relax
  \def\href#1#2{#2} \def\path#1{#1}\fi

\bibitem{guijarro2021}
L.~Guijarro, J.~R. Vidal, V.~Pla,
  \href{https://doi.org/10.1109/ACCESS.2021.3078562}{Competition between
  service providers with strategic resource allocation: Application to network
  slicing}, {IEEE} Access 9 (2021) 76503--76517.
\newblock \href {https://doi.org/10.1109/ACCESS.2021.3078562}
  {\path{doi:10.1109/ACCESS.2021.3078562}}.
\newline\urlprefix\url{https://doi.org/10.1109/ACCESS.2021.3078562}

\bibitem{caballero2017b}
P.~Caballero, A.~Banchs, G.~de~Veciana, X.~Costa-P{\'e}rez, Network slicing
  games: Enabling customization in multi-tenant networks, in: INFOCOM 2017-IEEE
  Conference on Computer Communications, IEEE, 2017, pp. 1--9.

\bibitem{sacoto2020}
E.~J. Sacoto-Cabrera, L.~Guijarro, J.~R. Vidal, V.~Pla, Economic feasibility of
  virtual operators in 5{G} via network slicing, Future Generation Computer
  Systems 109 (2020) 172--187.

\bibitem{shafiei2018}
S.~Shafiei, Z.~Gu, M.~Saberi,
  \href{https://www.sciencedirect.com/science/article/pii/S1569190X18300558}{Calibration
  and validation of a simulation-based dynamic traffic assignment model for a
  large-scale congested network}, Simulation Modelling Practice and Theory 86
  (2018) 169--186.
\newblock \href {https://doi.org/https://doi.org/10.1016/j.simpat.2018.04.006}
  {\path{doi:https://doi.org/10.1016/j.simpat.2018.04.006}}.
\newline\urlprefix\url{https://www.sciencedirect.com/science/article/pii/S1569190X18300558}

\bibitem{train2009}
K.~E. Train, Discrete choice methods with simulation, Cambridge university
  press, 2009.

\bibitem{mcfadden1974}
D.~McFadden, et~al., Conditional logit analysis of qualitative choice behavior.
  1973, Frontiers in Econometrics, ed. P. Zarembka (1974) 105--42.

\bibitem{guadagni1983}
P.~M. Guadagni, J.~D. Little, A logit model of brand choice calibrated on
  scanner data, Marketing science 2~(3) (1983) 203--238.

\bibitem{small2007}
K.~A. Small, E.~T. Verhoef, R.~Lindsey, The economics of urban transportation,
  Routledge, 2007.

\bibitem{meignan2007}
D.~Meignan, O.~Simonin, A.~Koukam,
  \href{https://www.sciencedirect.com/science/article/pii/S1569190X07000263}{Simulation
  and evaluation of urban bus-networks using a multiagent approach}, Simulation
  Modelling Practice and Theory 15~(6) (2007) 659--671.
\newblock \href {https://doi.org/https://doi.org/10.1016/j.simpat.2007.02.005}
  {\path{doi:https://doi.org/10.1016/j.simpat.2007.02.005}}.
\newline\urlprefix\url{https://www.sciencedirect.com/science/article/pii/S1569190X07000263}

\bibitem{benakiva1985}
M.~Ben-Akiva, S.~R. Lerman, Discrete choice analysis: theory and application to
  travel demand, Vol.~9, MIT press, 1985.

\bibitem{bortolomiol2022}
S.~Bortolomiol, V.~Lurkin, M.~Bierlaire, Price-based regulation of
  oligopolistic markets under discrete choice models of demand, Transportation
  49~(5) (2022) 1441--1463.

\bibitem{anderson2020}
S.~P. Anderson, N.~Erkal, D.~Piccinin, Aggregative games and oligopoly theory:
  short-run and long-run analysis, The RAND Journal of Economics 51~(2) (2020)
  470--495.

\bibitem{motta2021}
M.~Motta, E.~Tarantino, The effect of horizontal mergers, when firms compete in
  prices and investments, International Journal of Industrial Organization 78
  (2021) 102774.

\bibitem{rodini2003}
M.~Rodini, M.~R. Ward, G.~A. Woroch, Going mobile: substitutability between
  fixed and mobile access, Telecommunications Policy 27~(5-6) (2003) 457--476.

\bibitem{maicas2009}
J.~P. Maicas, Y.~Polo, F.~J. Sese, Reducing the level of switching costs in
  mobile communications: The case of mobile number portability,
  Telecommunications Policy 33~(9) (2009) 544--554.

\bibitem{maille2014}
P.~Maill{\'e}, B.~Tuffin, Telecommunication network economics: from theory to
  applications, Cambridge University Press, 2014.

\bibitem{coucheney2013}
P.~Coucheney, P.~Maill{\'e}, B.~Tuffin, Impact of competition between isps on
  the net neutrality debate, IEEE Transactions on Network and Service
  Management 10~(4) (2013) 425--433.

\bibitem{caron2010}
S.~Caron, G.~Kesidis, E.~Altman, Application neutrality and a paradox of side
  payments, in: Proceedings of the Re-Architecting the Internet Workshop, 2010,
  pp. 1--6.

\bibitem{shin2014}
J.~Shin, M.~Jo, J.~Lee, D.~Lee, Strategic management of cloud computing
  services: Focusing on consumer adoption behavior, IEEE Transactions on
  engineering management 61~(3) (2014) 419--427.

\bibitem{guijarro2017jsac}
L.~Guijarro, V.~Pla, J.~R. Vidal, M.~Naldi, Game theoretical analysis of
  service provision for the internet of things based on sensor virtualization,
  {IEEE} Journal on Selected Areas in Communications 35~(3) (2017) 691--706.

\bibitem{reichl2011}
P.~Reichl, B.~Tuffin, R.~Schatz, Logarithmic laws in service quality
  perception: where microeconomics meets psychophysics and quality of
  experience, Telecommunication Systems (2011) 1--14.

\bibitem{itu2009}
ITU-R, Guidelines for evaluation of radio interface technologies for
  {IMT}-advanced, Tech. Rep. M.2135-1, ITU (December 2009).

\end{thebibliography}
\end{document}